\documentclass[a4paper,11pt]{article}
\pdfoutput=1 

\usepackage{jheppub} 

\usepackage[T1]{fontenc} 

\usepackage{MnSymbol}

\newcommand{\floor}[1]{\lfloor #1 \rfloor}

\title{\boldmath Polynomial reduction and evaluation of tree- and loop-level CHY amplitudes}

\author[a]{Michael Zlotnikov}

\affiliation[a]{Brown University\\Department of Physics\\182 Hope St, Providence, RI, 02912}

\emailAdd{michael\_zlotnikov@brown.edu}

\abstract{\\We develop a polynomial reduction procedure that transforms any gauge fixed CHY amplitude integrand for $n$ scattering particles into a $\sigma$-moduli multivariate polynomial of what we call the \textit{standard form}. We show that a standard form polynomial must have a specific \textit{ladder type} monomial structure, which has finite size at any $n$, with highest multivariate degree given by $(n-3)(n-4)/2$. This set of monomials spans a complete basis for polynomials with rational coefficients in kinematic data on the support of scattering equations. Subsequently, at tree and one-loop level, we employ the global residue theorem to derive a prescription that evaluates any CHY amplitude by means of collecting simple residues at infinity only. The prescription is then applied explicitly to some tree and one-loop amplitude examples.}

\begin{document} 
\maketitle
\flushbottom

\section{Introduction}
After the amazing discovery of the relation between perturbative gauge theory and twistor string theory by Witten \cite{Witten:2003nn}, there have been several developments on computing scattering matrices in various theories from a moduli space on a punctured sphere \cite{Roiban:2004yf,Cachazo:2012da,Cachazo:2012kg,Huang:2012vt,Cachazo:2013iaa}. Cachazo, He and Yuan (CHY) proposed the equations governing the map from the space of kinematic invariants to the moduli space to be the same in each case and independent of the particular spacetime dimension. This led them to search for a more general formulation of scattering matrices in arbitrary dimension. Deriving some inspiration from a formula for MHV gravity amplitudes due to Hodges \cite{Bern:1998sv,Hodges:2012ym,Nguyen:2009jk}, CHY went on to discover their new formulation for amplitudes in a range of theories in \cite{Cachazo:2013gna,Cachazo:2013hca,Cachazo:2013iea}, and later \cite{Cachazo:2014nsa,Cachazo:2014xea}. This so called CHY formulation produces tree level $n$-point scattering amplitudes for massless particles in arbitrary dimension by means of $(n-3)$ moduli integrations localizing so called scattering equations. The scattering equations first appeared in the work of Fairlie and Roberts \cite{FairlieRoberts}, and later Gross and Mende \cite{GrossMende}, as well as more recently Witten \cite{Witten:2004cp}, and from the string theory classical worldsheet perspective in\footnote{The author thanks P. Caputa for pointing out this last point.} \cite{Caputa:2011zk,Caputa:2012pi}. Soon after the CHY equations made their appearance, the scalar and gluon cases were proven directly \cite{Dolan:2013isa} by means of BCFW recursion relations \cite{Britto:2004ap,Britto:2005fq}. Subsequently generalizations appeared, extending the formulation in terms of scattering equations to involve i.e. massive particles \cite{Dolan:2013isa,Naculich:2014naa,Naculich:2015zha,Naculich:2015coa}, fermions \cite{Weinzierl:2014ava}, supersymmetric theory \cite{Adamo:2015gia,Adamo:2015hoa}, one-loop amplitudes \cite{Adamo:2013tsa,Geyer:2015bja,Geyer:2015jch}, QCD related amplitudes \cite{delaCruz:2015raa}, off-shell amplitudes \cite{Lam:2015mgu}, or comparison to a string theory setting \cite{Mason:2013sva,Bjerrum-Bohr:2014qwa,Casali:2015vta}.\\
The most direct approach to evaluate amplitudes in CHY formulation was to try and find solutions to the scattering equations in general \cite{Weinzierl:2014vwa,Lam:2014tga}, or solve at special kinematics \cite{Kalousios:2013eca,Naculich:2014rta}. The scattering equations could also be reformulated in a polynomial form \cite{Dolan:2014ega,He:2014wua}. However, it became clear that solving scattering equations is very non-trivial and is not the most convenient way of evaluating amplitudes. Subsequently, techniques that avoid explicit solving of scattering equations started to emerge \cite{Kalousios:2015fya}. Contour deformations in the moduli integrals led to diagrammatic prescriptions that can be used to evaluate separate amplitude building blocks \cite{Cachazo:2015nwa,Baadsgaard:2015voa,Baadsgaard:2015ifa,Baadsgaard:2015hia,Lam:2015sqb}. An algebraic approach to evaluating scattering amplitudes in CHY formulation involving so-called companion matrices was suggested in \cite{Huang:2015yka}. For a comparison of this method with an elimination theory based technique see \cite{Cardona:2015eba}. One further algebraic technique involving polynomial inversion of moduli differences on the support of the ideal spanned by scattering equations, as well as the Bezoutian matrix to evaluate amplitudes was presented in \cite{Sogaard:2015dba}. Elimination theory was applied to scattering equations in polynomial form to obtain single variable polynomials \cite{Cardona:2015ouc,Dolan:2015iln}. Loop level integrands have been shown to follow from higher dimensional massless tree-level amplitudes \cite{Cachazo:2015aol,Feng:2016nrf}. Some further progress on evaluating CHY amplitudes was made in \cite{Lam:2016tlk}, diagrammatic techniques were generalized to compute higher order poles \cite{Huang:2016zzb}, and a double cover deformation of the moduli space led to evaluation of more general amplitude types as well \cite{Gomez:2016bmv}. Finally, monodromy relations were applied to Yang-Mills amplitudes in CHY representation to facilitate evaluation \cite{Bjerrum-Bohr:2016juj}.\\
In this work we start by developing a polynomial degree reduction procedure for multivariate polynomials in $\sigma$-moduli on the support of gauge fixed scattering equations for any $n$. As a consequence we realize that the most general multivariate polynomial in $\sigma$-moduli can be reduced to contain what we call \textit{ladder type} monomials only, with multivariate degree of at most $\frac{(n-3)(n-4)}{2}$ and coefficients rational in kinematic data. We say such a fully reduced polynomial is of \textit{standard form}. Application of Hilbert's strong Nullstellensatz as well as our degree reduction procedure conceptually allows us to find a standard form polynomial expression for rational functions in the $\sigma$-moduli. Making use of the above findings, a CHY amplitude integrand of any theory at any $n$ can be converted to a corresponding standard form polynomial. This general structural constraint is one of the main findings of the current work. After the polynomial reduction is carried out, we use the global residue theorem to derive a prescription to evaluate CHY amplitudes by collecting simple residues at infinity only. We note that only highest degree ladder type monomials contribute to any such amplitude integral, and since we find only simple poles the evaluation step is trivial. The difficulty is shifted towards finding standard form polynomial integrands for CHY amplitudes. We demonstrate the prescription on explicit examples of amplitude integrands at tree and one-loop level.\\
This paper is organized as follows. In section \ref{sec:CHY} we review the CHY formulation of tree-level scattering amplitudes for massless $\phi^3$ scalar theory as an example. As a warm up, section \ref{sec:warmup} shows a five point amplitude calculation to motivate our further investigation in section \ref{sec:reduction}. Section \ref{app:degred} describes the degree reduction of multivariate polynomials to the standard form, and section \ref{sec:polinv} extends the reduction procedure to rational functions, on the support of gauge fixed scattering equations. Subsequently, section \ref{sec:contours} describes the global residue theorem based proof for our amplitude evaluation prescription after polynomial reduction is applied to the integrands. In section \ref{sec:examples} we give explicit examples on how amplitudes are evaluated making use of our new method. We go on to consider 1-loop amplitudes in section \ref{sec:loop}, where we determine gauge fixed polynomial scattering equations that are free of singular solutions in the forward limit. Section \ref{sec:loopexpl} contains a few amplitude evaluation examples at 1-loop. We conclude in section \ref{sec:conclusion}. Appendix \ref{app:ratmom} suggests a simple method to generate real rational on-shell momenta based on Euclid's Pythagorean triple parametrization.  

\paragraph{Note added:}{$~$}\\
When this work was being prepared for submission, J. Bosma, M. S{\o}gaard and Y. Zhang released a paper with similar results in \cite{Bosma:2016ttj}.

\section{CHY formulation of tree level scattering amplitudes}
\label{sec:CHY}
The Cachazo-He-Yuan (CHY) formulation of tree-level scattering amplitudes for massless particles in arbitrary dimension was introduced in \cite{Cachazo:2013gna,Cachazo:2013hca}. In CHY representation, the map of kinematic data to the moduli space is governed by the rational scattering equations
\begin{align}
 \label{ratf}
f_a=\sum_{b=1,b\neq a}^n\frac{k_a\cdot k_b}{\sigma_a-\sigma_b}~~~~~~\forall a\in\{1,2,...,n\}.
\end{align}
Dolan and Goddard transformed the original amplitude expression to involve polynomial scattering equations \cite{Dolan:2014ega}. In what follows, it will be more convenient for us to work with polynomial scattering equations, therefore we will use the latter form for i.e. an $n$-point scalar $\phi^3$ amplitude in the examples to follow:
\begin{align}
 \label{CHYDG}
 A_n=\int \left(\prod_{{c=1}\atop{c\neq q,p,w}}^n d\sigma_c\right)(\sigma_{qp}\sigma_{pw}\sigma_{wq})\left(\prod_{1\leq i<j\leq n}\sigma_{ij}\right)\left(\prod_{a=2}^{n-2}\delta\left(\tilde{h}_a\right)\right)\frac{1}{(\sigma_{12}\sigma_{23}...\sigma_{n1})^{2}}.
\end{align}
Here the indices $1\leq q <p<w\leq n$ are fixed and can be chosen arbitrarily without changing the result. Minkowski momenta of scattering external particles are denoted $k_i$, and the difference of moduli is abbreviated as $\sigma_{ij}=\sigma_i-\sigma_j$. There are $n-3$ moduli integrations and the same amount of delta functions, such that the integral reduces to a sum over the solutions to the system of the scattering equations in the delta function arguments
\begin{align}
\label{scatOrig}
\tilde{h}_i\equiv \sum_{\{q_1,...,q_i\}\subset\{1,2,...,n\}}\mathfrak{s}_{q_1,...,q_i}\prod_{j=1}^i \sigma_{q_j}=0.
\end{align}
In this formula the summation is over all possible unordered subsets of $i$ different numbers $\{q_1,...,q_i\}$ out of the integer sequence from $1$ to $n$. Due to momentum conservation and massless on-shell conditions, the kinematic variables
\begin{align}
\label{kinS}
\mathfrak{s}_{q_1,...,q_i}=\frac{1}{2}\left(\sum_{j=1}^ik_{q_j}\right)^2
\end{align}
 are only non-zero when at least $2$ or at most $n-2$ indices are provided. Therefore, exactly $n-3$ scattering equations (\ref{scatOrig}) from $\tilde{h}_2$ through $\tilde{h}_{n-2}$ are nontrivial.\\
In the following we will be working with the particular gauge choice $\sigma_1=\infty$, $\sigma_2=0$ and $\sigma_3=1$ for convenience. For this purpose we define the gauge fixed polynomial scattering equations:
\begin{align}
\label{scat}
 h_i\equiv\left(\lim_{\sigma_1\rightarrow \infty}\frac{1}{\sigma_1}\tilde{h}_{i+1}\right)|_{{\sigma_2=0}\atop{\sigma_3=1}}=0~~~,~~~\forall i\in\{1,2,...,n-3\}.
\end{align}
Correspondingly, we will fix the free indices in (\ref{CHYDG}) as $q=1,p=2,w=3$.

\section{Warm up: five point tree level scalar amplitude}
\label{sec:warmup}
At five points we have two scattering equations:
\begin{align}
 h_1=&\sigma _4 \mathfrak{s}_{1,4}+\sigma _5 \mathfrak{s}_{1,5}+\mathfrak{s}_{1,3}=0,\notag\\
 h_2=&\sigma _4 \sigma _5 \mathfrak{s}_{2,3}+\sigma _5 \mathfrak{s}_{2,4}+\sigma _4 \mathfrak{s}_{2,5}=0.\notag
\end{align}
The gauge fixed scattering amplitude for scalars becomes
\begin{align}
\label{eq:ampl5sc}
A^{\phi^3}_5=\oint \frac{d\sigma_4d\sigma_5}{ h_1 h_2}\frac{\sigma _4  \sigma _5\left(1-\sigma _5\right)}{\left(1-\sigma _4\right) \left(\sigma _4-\sigma
   _5\right)},
\end{align}
where the delta functions have been mapped to simple poles as usual, and the integration contour is such that both poles are localized. We would like to transform the integrand such that an evaluation via contour deformation becomes simpler. For that end, consider the following equality
\begin{align}
\label{5ptansatz}
	\sigma _4  \sigma _5\left(1-\sigma _5\right)\hateq \left(1-\sigma _4\right) \left(\sigma _4-\sigma_5\right)N^{\phi^3}_5
\end{align}
where $\hateq$ shall denote equivalence on the support of scattering equations. Here $N^{\phi^3}_5$ clearly corresponds to the explicit integrand part of (\ref{eq:ampl5sc}). We claim that (\ref{5ptansatz}) can be realized i.e. by the following Ansatz
\begin{align}
\label{5ptNansatz}
	N^{\phi^3}_5=c_1 \sigma_4+c_2\sigma_5.
\end{align}
To show that this is indeed the case, we can first solve $ h_1=0$ for either $\sigma_4$ or $\sigma_5$, and solve $ h_2=0$ for $\sigma_4\sigma_5$:
\begin{align}
\label{5ptsc1s}
\sigma _4 =-\frac{ \mathfrak{s}_{1,5}}{\mathfrak{s}_{1,4}}&\sigma _5-\frac{\mathfrak{s}_{1,3}}{\mathfrak{s}_{1,4}},~~~~~~\sigma _5=-\frac{\mathfrak{s}_{1,4}}{\mathfrak{s}_{1,5}}\sigma _4  -\frac{\mathfrak{s}_{1,3}}{\mathfrak{s}_{1,5}},\\
\label{5ptsc2s}
&\sigma _4 \sigma _5= - \frac{\mathfrak{s}_{2,4}}{\mathfrak{s}_{2,3}}\sigma _5- \frac{\mathfrak{s}_{2,5}}{\mathfrak{s}_{2,3}}\sigma _4.
\end{align}
Then we start with (\ref{5ptansatz}) making use of (\ref{5ptNansatz}), expand both sides of the equation, and iterate the following substitution rules:
\begin{enumerate}
\item Whenever we encounter a monomial featuring both $\sigma_4$ and $\sigma_5$, we isolate the highest power of $\sigma_4\sigma_5$, substitute in the right hand side of (\ref{5ptsc2s}) and expand - this leads to an overall multivariate degree reduction in monomials.
\item Whenever we encounter a monomial featuring $\sigma_4$ xor $\sigma_5$ to a power higher than one, we isolate a single power of $\sigma_4$ xor $\sigma_5$ respectively, substitute it by the right hand side of the respective equation in (\ref{5ptsc1s}) and expand - this leads either to an overall degree reduction in monomials, or to creation of new $\sigma_4\sigma_5$ terms.
\end{enumerate}
Iterating the above two steps a few times reduces both sides of (\ref{5ptansatz}) to only the two monomials $\sigma_4$ and $\sigma_5$ with some constant coefficients.\footnote{The exact coefficients are not necessarily unique and might depend on the order of substitutions during the reduction.} Collecting all terms on one side of the equation and demanding that the overall coefficients of monomials $\sigma_4$ and $\sigma_5$ vanish identically, we obtain a set of two linear equations in two unknowns $c_1$ and $c_2$. Solving these equations yields one possible solution for the Ansatz $N^{\phi^3}_5$, i.e.
\begin{align}
	c_1=&\frac{\mathfrak{s}_{1,4}\mathfrak{s}_{2,5} \left(\left(\mathfrak{s}_{1,3}+\mathfrak{s}_{1,5}\right)
   \mathfrak{s}_{2,4}+\mathfrak{s}_{1,4} \left(\mathfrak{s}_{2,3}+2
   \mathfrak{s}_{2,4}+\mathfrak{s}_{2,5}\right)\right)}{
   \left(\left(\mathfrak{s}_{1,3}+\mathfrak{s}_{1,4}\right)
   \left(\mathfrak{s}_{2,3}+\mathfrak{s}_{2,4}\right)-\mathfrak{s}_{1,5}
   \mathfrak{s}_{2,5}\right) \left(\mathfrak{s}_{1,3}
   \mathfrak{s}_{2,3}-\left(\mathfrak{s}_{1,4}+\mathfrak{s}_{1,5}\right)
   \left(\mathfrak{s}_{2,4}+\mathfrak{s}_{2,5}\right)\right)},\notag\\
	c_2=&\frac{\mathfrak{s}_{2,4}(\mathfrak{s}_{1,3}+\mathfrak{s}_{1,4})}{\mathfrak{s}_{1,5}
   \mathfrak{s}_{2,5}-\left(\mathfrak{s}_{1,3}+\mathfrak{s}_{1,4}
   \right) \left(\mathfrak{s}_{2,3}+\mathfrak{s}_{2,4}\right)}+\frac{\mathfrak{s}_{1,5}\mathfrak{s}_{2,4}}{\left(\mathfrak{s}_{1,4}+\mathfrak{s}_{1,5}\right)
   \left(\mathfrak{s}_{2,4}+\mathfrak{s}_{2,5}\right)-\mathfrak{s}_{1,3}
   \mathfrak{s}_{2,3}}.\notag
\end{align}
Deforming the integration contours to infinity consecutively, we find only simple poles and get \footnote{In what follows, we give more details on this, from the point of view of global residue theorem.}
\begin{align}
\label{eq:ampl5scRes}
A^{\phi^3}_5=&\oint d\sigma_4 d\sigma_5\frac{N^{\phi^3}_5}{ h_1 h_2}=\frac{c_1}{\mathfrak{s}_{1,4} \mathfrak{s}_{2,3}}-\frac{c_2}{\mathfrak{s}_{1,5}
   \mathfrak{s}_{2,3}}.
\end{align}
Using momentum conservation and the fact that all external particles are massless, we can re-express the above in the following familiar form 
\begin{align}
A^{\phi^3}_5=\frac{1}{\mathfrak{s}_{1,2}
   \mathfrak{s}_{3,4}}+\frac{1}{\mathfrak{s}_{5,1}\mathfrak{s}_{2,3} }+\frac{1}{\mathfrak{s}_{4,5}\mathfrak{s}_{1,2}}+\frac{1}{\mathfrak{s}_{3,4}\mathfrak{s}_{5,1} }+\frac{1}{\mathfrak{s}_{2,3}
   \mathfrak{s}_{4,5}},
	\end{align}
confirming that the result we found is indeed the correct five point massless scalar amplitude in  $\phi^3$ theory. In the following section we will generalize the above technique to all $n$.

\section{Amplitude structure and evaluation prescription}
\label{sec:reduction}
Our plan is to show that any multivariate polynomial on the support of scattering equations can be written in a specific monomial structure we call the \textit{standard form}. Subsequently, we show that any rational function that is finite and non-vanishing on the support of scattering equations can be written as a standard form polynomial. Lastly, we apply these findings to amplitude integrands, convert them to standard form polynomials and evaluate the amplitude by means of the global residue theorem while collecting simple pole residues at infinity only.
\subsection{Degree reduction of polynomials to a standard form}
\label{app:degred}
In this section we start with an arbitrary multivariate polynomial $N$ in the $n-3$ different $\sigma$-moduli that are not gauge fixed (substitute $n\rightarrow n+2$ everywhere for 1-loop), and show that any such polynomial can be degree reduced to a very specific form.
\paragraph{Conventions:} Consider a generic monomial $M$ within polynomial $N$ separately:
\begin{align}
\label{genmonom}
M=C \sigma_{q_{1}}^{p_{1}}\sigma_{q_{2}}^{p_{2}}...\sigma_{q_{m_{max}}}^{p_{m_{max}}}.
\end{align}
$C$ is an overall constant, $q_{1}\neq q_{2}\neq...\neq q_{m_{max}}$ label the different $\sigma$-moduli appearing in the monomial $M$, while $p_{1},p_{2},...,p_{m_{max}}$ are the corresponding powers of each $\sigma$-modulus. We choose to always order all $\sigma$-moduli within each monomial such that $p_{1}\leq p_{2}\leq...\leq p_{m_{max}}$. For convenience we define $p_{0}\equiv 0$ for all $M$. Since there are at most $n-3$ different non-gauge fixed $\sigma$-moduli, we have $0\leq m_{max}\leq n-3$ in general.\footnote{The case $m_{max}=0$ corresponds to only $C$ being present in (\ref{genmonom}).}
\paragraph{Definition 1:} We define a monomial $M$ as introduced in the conventions above to be of \textit{ladder type} if its moduli powers satisfy $0\leq p_{j}-p_{j-1}\leq 1$ for all $j\in\{1,2,...,m_{max}\}$ when $m_{max}>0$, and iff additionally the property $0\leq m_{max}\leq n-4$ is satisfied.
\paragraph{Definition 2:} We define a multivariate polynomial in the non-gauge fixed $\sigma$-moduli to be of \textit{standard form} if it consists of \textit{ladder type} monomials only, with coefficients rational in kinematic data. See Table \ref{tab:ladmon} for some examples of ladder type monomials.
\begin{table}[h]
\centering
  \begin{tabular}{ | c | c | c |}
    \hline
    $n=4$ & $n=5$ & $n=6$ \\ \hline
    $1$ & $1,~\sigma_4,~\sigma_5$ & $1,~\sigma_4,~\sigma_5,~\sigma_6,~\sigma_4\sigma_5,~\sigma_4\sigma_6,~\sigma_5\sigma_6,~\sigma_5\sigma_4^2,~\sigma_6\sigma_4^2,~\sigma_6\sigma_5^2,~\sigma_4\sigma_5^2,~\sigma_4\sigma_6^2,~\sigma_5\sigma_6^2$ \\ \hline
  \end{tabular}
\caption{Examples of all ladder type monomials for the first few $n$. ($\sigma_1,\sigma_2,\sigma_3$ gauge fixed.)}  
\label{tab:ladmon}
\end{table}
\paragraph{Theorem 1:}\textit{On the support of the ideal spanned by scattering equations, an arbitrary regular multivariate polynomial $N$ in the $n-3$ non-gauge fixed moduli, with coefficients rational in kinematic data, is equivalent to at least one }standard form \textit{polynomial $N'$ that consists of ladder type monomials only, with coefficients rational in kinematic data.}
\paragraph{Proof:} To prove this we use flow arguments induced by scattering equation based transformations in the space of moduli powers within monomials. The arguments consist of the following two steps.
\paragraph{Step 1: Reduction of monomials to $0\leq p_{j}-p_{j-1}\leq 1$ for all $j\in\{1,2,...,m_{max}\}$}{$~$}\\
Consider a generic monomial of an arbitrary polynomial 
\begin{align}
\label{arbimonom}
C \underbrace{\sigma_{q_{1}}^{p_{1}}\sigma_{q_{2}}^{p_{2}}...\sigma_{q_{j-1}}^{p_{j-1}}}_{w_1\text{ terms}}\underbrace{\sigma_{q_{j}}^{p_{j}}...\sigma_{q_{m_{max}}}^{p_{m_{max}}}}_{w_2\text{ terms}},
\end{align}
for some fixed $1\leq j\leq m_{max}$, ordered as $p_{1}\leq p_{2}\leq...\leq p_{m_{max}}$ and such that $p_{j}-p_{j-1}>1$, so that the monomial is non-ladder type. Also note that $0\leq (w_1+w_2=m_{max})\leq n-3$. If we want to transform this monomial into a sum over ladder type monomials, we first have to reduce the discrepancy $p_{j}-p_{j-1}>1$ to $0\leq p_{j}-p_{j-1}\leq 1$. We employ the scattering equations to do that as follows.\\
The general structure of gauge fixed polynomial scattering equations $ h_a=0$ for $a=1,...,n-3$ is such that $ h_a$ features all possible multilinear monomials of degree $a$ and $a-1$ respectively. Therefore, we can solve the scattering equation $ h_{w_2}=0$ for the monomial $\underbrace{\sigma_{q_{j}}...\sigma_{q_{m_{max}}}}_{w_2\text{ terms}}$:
\begin{align}
\label{Pfunct}
\sigma_{q_{j}}...\sigma_{q_{m_{max}}}=\sigma_{q_{j}}...\sigma_{q_{m_{max}}}-\frac{ h_{w_2}}{(\partial_{\sigma_{q_{j}}}...\partial_{\sigma_{q_{m_{max}}}} h_{w_2})}.
\end{align}
The derivatives in the denominator isolate the coefficient of monomial $\sigma_{q_{j}}...\sigma_{q_{m_{max}}}$ within $ h_{w_2}$. This coefficient is canceled for the corresponding summand in the numerator and the pure monomial is subtracted. Therefore, the right hand side of (\ref{Pfunct}) features all possible multilinear monomials of degree $w_2-1$ and all multilinear monomials of degree $w_2$ except for $\sigma_{q_{j}}...\sigma_{q_{m_{max}}}$.\\
We can now isolate $(\sigma_{q_{j}}...\sigma_{q_{m_{max}}})^{\floor{\frac{p_{j}-p_{j-1}}{2}}}$ moduli from the $w_2$ terms in (\ref{arbimonom}), substitute them by the right hand side of (\ref{Pfunct}) to the power $\floor{\frac{p_{j}-p_{j-1}}{2}}$ and expand.\footnote{The notation $\floor{x}$ means the floor function, returning the biggest integer $\leq x$.} Since each multilinear monomial of a certain degree is unique up to a constant factor, this has the effect that in each of the resulting terms 
 \begin{itemize} 
        \item the power of at least one modulus in the $w_2$ terms is reduced by at least one,
        \item the power of at least one modulus in the $w_1$ terms is increased by at least one\footnote{Note that the power of this modulus could have been zero initially.}, or the overall degree is reduced. 
     \end{itemize}
Since the above guarantees a non-zero flow in the distribution of $\sigma$-moduli powers away from $w_2$ terms either into the $w_1$ terms or into overall degree reduction, iteration of the substitution rule for all $j$ and each monomial in the resulting terms is bound to reach a fixed point. This fixed point is straightforwardly given by the state where all monomials obey $0\leq p_{j}-p_{j-1}\leq 1$ for all $j\in\{1,2,...\}$ within each respective monomial, since then $\floor{\frac{p_{j}-p_{j-1}}{2}}=0$ for all $j$ and no substitutions can be carried out any more.

\paragraph{Step 2: Reduction of monomials to $m_{max}\leq n-4$}{$~$}\\
After step 1 is applied to all monomials in a polynomial $N$, it can still contain monomials with the maximal number of different moduli $m_{max}=n-3$:
\begin{align}
\label{almostladder}
C \sigma_{q_{1}}^{p_{1}}\sigma_{q_{2}}^{p_{2}}...\sigma_{q_{n-3}}^{p_{n-3}},
\end{align}
with $p_{1}=1$ and $0\leq p_{j}-p_{j-1}\leq 1$ for all $j\in\{2,3,...,n-3\}$. Similar to (\ref{Pfunct}), we can solve the gauge fixed polynomial scattering equation $ h_{n-3}=0$ for the single highest degree multilinear monomial $\sigma_{q_{1}}\sigma_{q_{2}}...\sigma_{q_{n-3}}$. Since that yields only multilinear terms of degree $n-4$, this necessarily leads to a degree reduction. We isolate the highest power of $\sigma_{q_{1}}\sigma_{q_{2}}...\sigma_{q_{n-3}}$ from monomials such as (\ref{almostladder}), make the substitution obtained from $ h_{n-3}=0$ and expand. Due to the guaranteed degree reduction in this step, we are again bound to iteratively reach a fixed point. This fixed point is trivially given by the condition $m_{max}\leq n-4$ for all resulting monomials $M$, since then no highest degree multilinear monomial can be isolated within the monomials, and therefore no substitutions can be carried out any more.
\paragraph{Conclusion}{$~$}\\
Step 1 and 2 above can be applied consecutively and iteratively to an arbitrary multivariate polynomial $N$. Due to the guaranteed degree reduction in step 2, both fixed points are bound to be reached simultaneously eventually. Therefore, we have shown that any polynomial $N$ on the support of scattering equations can be cast into a \textit{standard form} $N'$ containing only ladder type monomials.\footnote{The coefficients stay rational in kinematic data since we only used a finite number of additions and multiplications, and the coefficients in the scattering equations are rational as well.} Note that the degrees of the ladder type monomials $M_{lt}$ are $0\leq \text{deg}(M_{lt}) \leq\frac{(n-3)(n-4)}{2}$ at $n$ points. The full set of pure ladder type monomials at any $n$ is symmetric in all moduli. This homogeneity follows from the homogeneity of scattering equations that are used to achieve this form.

\subsection{Polynomial reduction of rational expressions}
\label{sec:polinv}
\paragraph{Theorem 2:}\textit{On the support of the ideal spanned by the scattering equations, any regular\footnote{By regular we mean non-infinite and non-zero on all solutions to the scattering equations.} multivariate rational function $\frac{P}{Q}$ in the $n-3$ non-gauge fixed moduli, where $P$ and $Q$ are polynomials with rational coefficients in kinematic data, is equivalent to at least one }standard form \textit{polynomial $N'$ that consists of ladder type monomials only, with rational coefficients in kinematic data.}
\paragraph{Proof:} Similar to some ideas of \cite{Sogaard:2015dba}, we will make use of Hilbert's Nullstellensatz. Consider the following equation involving the set of gauge fixed polynomial scattering equations $ h_m$ and multivariate polynomials in the $\sigma$-moduli $P,Q,a$ and $a_m$ for $m\in\{1,2,...,n-3\}$
\begin{align}
\label{Shilbert}
 a\,Q+\sum_{m=1}^{n-3}{a_m  h_m}=P.
\end{align}
The strong version of the Nullstellensatz guarantees that we can always find polynomials $a$ and $a_m$ for given polynomials $P$ and $Q$ such that (\ref{Shilbert}) is satisfied, as long as the $a,a_m,P$ and $Q$ do not share common roots among themselves and with the set of scattering equation polynomials $ h_m$. Considering the situation at the locus of solutions to the scattering equations, this simplifies to
\begin{align}
\label{Shilbert2}
 a\,Q\hateq P,
\end{align}
where we use the symbol $\hateq$ to denote equivalence modulo the ideal spanned by the scattering equations. Thus, $a\hateq\frac{P}{Q}$ is a polynomial expression for a rational function.\footnote{Dividing by $Q$ is allowed since it is per assumption non-zero at the locus of solutions to the scattering equations.} Due to Theorem 1, a standard form polynomial $N'\hateq a$ must exist, which concludes the proof.
\paragraph{Construction:} In the above proof we used the fact that a standard form polynomial $N'\hateq a$ must exist, however the proof was not constructive. To construct an explicit $N'$ corresponding to a given rational function $\frac{P}{Q}$ we have to work harder. In principle, this step could be realized by various techniques. Here we will make use of an ad hoc procedure as follows.\\
Since the ladder type monomials span a complete polynomial basis with rational coefficients on the support of scattering equations, we can make an ansatz $\tilde{N}'$ containing all ladder type monomials with unfixed coefficients to parametrize our ignorance of what $N'$ actually is: $\tilde{N}'Q-P\hateq 0$. Making use of an implementation of the degree reduction procedure for Theorem 1, we can find a standard form polynomial $H'$ such that $H'\hateq \tilde{N}'Q-P\hateq0$. Demanding that the overall coefficient of each monomial in $H'$ vanishes separately, sets up a number of linear equations in (at least) the same number of unknown coefficients of $\tilde{N}'$.\footnote{Since there is a finite number of ladder type monomials at any $n$, the number of unfixed coefficients in $\tilde{N}'$ is at least equal to the number of monomials in a most general resulting standard form $H'$. If $H'$ has fewer than the maximum number of monomials, then the amount of unfixed coefficients is greater than the number of linear equations.} Solving this set of equations fixes the coefficients and yields an $N'$.
\\
In practice, in many cases of interest the ansatz for $\tilde N'$ does not require all ladder type monomials to be present to find a valid standard form polynomial $N'$. This reduces the dimension and complexity of the linear set of equations one has to solve. Additionally, we will see in the next section that only the coefficients of the highest degree ladder type monomials have a non-vanishing contribution to an amplitude integral.

\subsection{Collecting residues}
\label{sec:contours}
In this section we concentrate on tree level amplitudes for concreteness. However, at every step in the following it should be clear that essentially the same logic applies to the loop level integrands. Therefore, the result we find is valid in general.
\paragraph{Theorem 3:} 
\textit{Any amplitude integral of the general shape}\footnote{Here, again, we consider the formulation where the delta functions have been mapped to simple poles with appropriate integration contours. Factors of $2\pi i$ are suppressed.}
\begin{align}
\label{contamp}
A_n=\oint\frac{N(\sigma_4,\sigma_5,...,\sigma_n)}{\prod_{j=1}^{n-3} h_j}\prod_{i=4}^nd\sigma_i ,
\end{align}
\textit{where $h_j$ for $j=1,...,n-3$ are gauge fixed scattering equation polynomials, $N(\sigma_4,\sigma_5,...,\sigma_n)$ is a} standard form \textit{polynomial in the $n-3$ non-gauge fixed moduli and the integration contour is initially localized at the locus of scattering equation solutions, can be evaluated by the following anti-symmetrized sum over the $(n-3)!$ different orders of consecutive infinity residues\footnote{The square brackets in $\sigma_{[n}=\infty,~...,~\sigma_{4]}=\infty$ denote anti-symmetrization with respect to the moduli indices, so that i.e. $\text{Res}_{\sigma_{[5}=\infty,~\sigma_{4]}=\infty}=\frac{1}{2!}(\text{Res}_{\sigma_{5}=\infty}\text{Res}_{\sigma_{4}=\infty}-\text{Res}_{\sigma_{4}=\infty}\text{Res}_{\sigma_{5}=\infty})$. The right most residue operation always acts first.}}
\begin{align}
\label{finalA}
A_n=(-1)^{n-3} (n-3)!~\text{Res}_{\sigma_{[n}=\infty,~...,~\sigma_{5}=\infty,~\sigma_{4]}=\infty}\left[\frac{N}{\prod_{j=1}^{n-3} h_j}\right].
\end{align}
\paragraph{Note:} Instead of calculating the $(n-3)!$ residues to evaluate the amplitude integral, it is possible to employ an integrand deformation in which the $h_i$'s are replaced by their leading homogeneous parts lt$(h_i)$. With this, the sum over residues equals one single residue at the origin by the transformation law of multivariate residues. This is an efficient alternative approach \cite{Cattani:1994}.\footnote{The author thanks the JHEP referee for pointing this out.}
\paragraph{Proof of Theorem 3:}  Starting with (\ref{contamp}), it is straightforward to realize that any contour deformation away from the locus defined by the solutions to the scattering equations can possibly yield other residues only at infinity.\\
Decompose the numerator polynomial of the integrand into monomials $N=\sum_i N_i$. By additivity of integrals, consider the contour integral in pieces involving just one monomial $N_q=M\propto \prod_{r=4}^n\sigma_r^{a_{r}}$ at a time, where the integer powers $a_{r}\geq 0$ are such that $M$ is a ladder type monomial. Planning to investigate residues at infinity, we perform the substitution $\sigma_i\rightarrow1/\sigma_i$ and $d\sigma_i\rightarrow-d\sigma_i/\sigma_i^2$ for $i\in\{4,5,...,n\}$, to focus on residues at zero instead, so that
\begin{align}
\label{int1}
\oint\frac{\prod_{r=4}^n\sigma_r^{a_{r}}}{\prod_{j=1}^{n-3} h_j}\prod_{i=4}^nd\sigma_i\rightarrow \oint\frac{(-1)^{n-3}}{\left(\prod_{j=1}^{n-3}\hat{ h}_j\right)\left(\prod_{r=4}^n\sigma_r^{a'_{r}}\right)}\prod_{i=4}^nd\sigma_i,
\end{align}
where $a'_{r}=(a_{r}-n+5)$ is an abbreviation for the new integer exponents, and $\hat{ h}_j$ can be conveniently obtained from the gauge fixed scattering equations in the slightly different gauge $\sigma_1=0,~\sigma_2=\infty,~\sigma_3=1$.\footnote{The set of scattering equations is invariant under simultaneous inversion $\sigma\rightarrow1/\sigma$ of all $\sigma$-moduli (up to overall $\sigma$-moduli factors that here are accounted for by the powers $a'_{r}$), as long as we also invert the values of the gauge fixed moduli.}\\
Next we apply the global residue theorem (GRT), as for instance described in detail in \cite{ArkaniHamed:2009dn}. Consider a contour integral in $n-3$ variables over an integrand $1/f_1f_2...f_{n-3}$, such that the contours localize all possible poles in the integrand $f_i=0,\forall i$. Since all possible residues are collected in this way, it follows from the GRT that the result must be zero:
\begin{align}
\label{GRT}
\text{Res}_{\{f_1,f_2,...,f_{n-4},f_{n-3}\}}=0.
\end{align}
Using the above in our integrand of interest in (\ref{int1}), assign $f_i=\hat{ h}_i$ for $i=\{1,2,...,n-4\}$ and $f_{n-3}=\hat{ h}_{n-3}\prod_{r=4}^n{\sigma_r}^{a'_{r}}$. This clearly takes all possible poles into consideration, so that eq. (\ref{GRT}) is satisfied. Expand the global residue as a sum over the poles in $f_{n-3}$:
\begin{align}
\label{GRTdecomp}
\text{Res}_{\{f_1,f_2,...,f_{n-4},f_{n-3}\}}=\text{Res}_{\{f_1,f_2,...,f_{n-4},\hat{ h}_{n-3}\}}+\sum_{t=4}^{n}\text{Res}_{\{f_1,f_2,...,f_{n-4},{\sigma_t}^{a'_{t}}\}}=0.
\end{align}
The first summand corresponds to (\ref{int1}), so that we can re-express it in terms of the other $n-3$ residues $\text{Res}_{\{f_1,f_2,...,f_{n-4},\hat{ h}_{n-3}\}}=-\sum_{t=4}^{n}\text{Res}_{\{f_1,f_2,...,f_{n-4},{\sigma_t}^{a'_{t}}\}}$. Whenever partial poles in a multivariate residue calculation depend on one variable only, single variable complex analysis can be used to integrate out the corresponding residue separately. In our case each $\text{Res}_{\{f_1,f_2,...,f_{n-4},{\sigma_t}^{a'_{t}}\}}$, among other poles, involves a pole $1/{\sigma_t}^{a'_{t}}$ dependent on a single variable $\sigma_t$, which we will now integrate out separately.\\
Considering that $a'_{t}=(a_{t}-n+5)$ for each $t$, only highest degree ladder type monomials have a non-vanishing contribution to the integral, since exactly one of their $a_{t}$ satisfies $a_{t}=n-4$ which produces a simple pole as $1/{\sigma_t}^{a'_{t}}$. For all other ladder type monomials we have $0\leq a_{t}<n-4$ such that $a'_{t}\leq 0$ and $1/{\sigma_t}^{a'_{t}}$ ceases to be a pole and thus no residue is present.\\
To keep track of the correct contour orientation in the remaining variables, we anti-commute $d\sigma_t$ to one side in the integration measure $d\sigma_4\wedge d\sigma_5\wedge ...\wedge d\sigma_n=(\pm)_t d\sigma_t \prod^n_{i=4,i\neq t}(\wedge d\sigma_i)$. This produces an overall plus or minus $(\pm)_t$ dependent on the initial position $t$. Thus, we have
\begin{align}
\label{step1}
 \oint\frac{\prod_{i=4}^nd\sigma_i}{\left(\prod_{j=1}^{n-3}\hat{ h}_j\right)\left(\prod_{r=4}^n\sigma_r^{a'_{r}}\right)}=\left\{ \begin{matrix}
- \oint\sum_{t=4}^n(\pm)_t\left(\frac{\prod_{{i=4}\atop{i\neq t}}^nd\sigma_i}{\left(\prod_{j=1}^{n-3}\hat{ h}_j\right)\left(\prod_{{r=4}\atop{r\neq t}}^n\sigma_r^{a'_{r}}\right)}\right)_{\sigma_t=0} & \text{ for }a'_{t}=1 \\
&\\
0 & \text{ for }a'_{t}<1 \end{matrix}, \right.
\end{align}
with the saturation $a'_{t}=1$ occurring for exactly one of the moduli in each highest degree ladder type monomial (nevertheless, we sum over all $\sum_{t=4}^n$ since it is not known a priori which label $t$ is going to yield the contribution).\footnote{
In terms of the expression in original variables on the left hand side of (\ref{int1}), this structurally means
\begin{align}
\label{int2}
\oint\frac{\prod_{r=4}^n\sigma_r^{a_{r}}}{\prod_{j=1}^{n-3} h_j}\prod_{i=4}^nd\sigma_i=&- \oint\sum_{u=4}^n(\pm)_u \text{Res}_{\sigma_u=\infty}\left[\frac{\prod_{r=4}^n\sigma_r^{a_{r}}}{\prod_{j=1}^{n-3} h_j}\right]\prod_{{i=4}\atop{i\neq u}}^nd\sigma_i,
\end{align}
where we imply that there are at most first order poles at infinity.}\\
As $\sigma_t=0$ in (\ref{step1}) is set, we find that $\hat{ h}_{n-3}$ reduces to a single monomial $\hat{ h}_{n-3}|_{\sigma_t=0}\propto \prod_{{j=4}\atop{j\neq t}}^n\sigma_j$ by general structure of scattering equations. Therefore, the non-vanishing contribution schematically becomes
\begin{align}
\label{int3}
\oint\sum_{t=4}^n(\pm)_t&\left(\frac{\prod_{{i=4}\atop{i\neq t}}^nd\sigma_i}{\left(\prod_{j=1}^{n-3}\hat{ h}_j\right)\left(\prod_{{r=4}\atop{r\neq t}}^n\sigma_r^{a'_{r}}\right)}\right)_{\sigma_t=0}= \oint\sum_{t=4}^n(\pm)_t\frac{C_{t}\prod_{{i=4}\atop{i\neq t}}^nd\sigma_i}{\left(\prod_{j=1}^{n-4}(\hat{ h}_j|_{\sigma_t=0})\right)\left(\prod_{{r=4}\atop{r\neq t}}^n\sigma_r^{a''_{r,t}}\right)}.
\end{align}
Here $C_{t}$ is one over the constant coefficient of the single monomial that survives as we take $\hat{ h}_{n-3}|_{\sigma_t=0}\propto \prod^n_{j=4,j\neq t}\sigma_j$, while the moduli of this monomial are accounted for by the new powers ${a''}_{r,t}$. The remaining $n-4$ scattering equation denominators $\hat{ h}_{j}|_{\sigma_t=0}$ now have the same monomial structure as scattering equation polynomials at $n-1$ points. Therefore, we can treat each summand in the sum over $t$ in (\ref{int3}) the same way as the initial expression (\ref{int1}), except now there is one fewer contour to integrate in each case. Thus, we can iterate. Noticing that by general structure of polynomial scattering equations we always get single monomials as more and more $\sigma_i$ are set to zero:
\begin{align}
\hat{ h}_{n-3}|_{\sigma_t=0}\propto \prod_{{j=4}\atop{j\neq t}}^n\sigma_j,~~~~~\hat{ h}_{n-4}|_{\sigma_t=0,~\sigma_l=0}\propto \prod_{{j=4}\atop{j\neq t,l}}^n\sigma_j,~~~~~\hat{ h}_{n-5}|_{\sigma_t=0,~\sigma_l=0,~\sigma_c=0}\propto \prod_{{j=4}\atop{j\neq t,l,c}}^n\sigma_j,~~~~~\text{etc.}\notag
\end{align}
ensures that each time a residue in a $\sigma$-modulus is collected, the remaining set of non-trivial scattering equation polynomials in the denominators is effectively reduced by one, as one of the scattering equation polynomials reduces to a single monomial and produces simple poles for the next iteration. With this, the above steps may be iterated from (\ref{int1}) to (\ref{int3}) $n-3$ times, while always expanding the resulting terms and summing over the process applied to one term at a time. Formally, each iteration adds one more level of signed infinity residue operations to (\ref{int2}). At the end of the day, when all contours have been treated, we are left with an anti-symmetrized sum over consecutive residue operations
\begin{align}
&\oint\frac{\prod_{r=4}^n\sigma_r^{a_{r}}}{\prod_{j=1}^{n-3} h_j}\prod_{i=4}^nd\sigma_i = (-1)^{n-3} (n-3)!~\text{Res}_{\sigma_{[n}=\infty,~...,~\sigma_{5}=\infty,~\sigma_{4]}=\infty}\left[\frac{\prod_{r=4}^n\sigma_r^{a_{r}}}{\prod_{j=1}^{n-3} h_j}\right].
\end{align}
This straightforwardly yields the full amplitude as we sum over all numerator monomials in the integrand, so that our final result for the amplitude is (\ref{finalA}). This concludes the proof.\\
\indent Due to the structure of standard form polynomials on the support of scattering equations we could rely on the fact that all residues we collect come from simple poles only. However, a straightforward generalization of the above steps yields the same result (\ref{finalA}) even for cases where $N$ is not a standard form polynomial and higher order residues are present.\\
It is interesting to note that the above procedure replaces a summation over $(n-3)!$ scattering equation solutions by a summation over the $(n-3)!$ different $(n-3)$-fold consecutive infinity residues in the $\sigma$-moduli. When $N$ is a standard form polynomial, all residues come from simple poles, such that the map from the integrand to the final result is trivial. With this the difficulty of the problem is shifted towards finding a standard form polynomial numerator $N$. Applying the degree reduction procedure described in the previous section this corresponds to solving a linear set on the order of $(n-3)!$ equations.

\subsection{Tree level amplitude examples}
\label{sec:examples}
In the following we demonstrate the evaluation prescription (\ref{finalA}) on $\phi^3$ scalar amplitudes at tree level. We also consider specific examples that otherwise require the more advanced evaluation techniques in order to be solved.
				\subsubsection{Six point tree level scalar example}
				At six points the three scattering equations are given by:
\begin{align}
 h_1&=\sigma _4 \mathfrak{s}_{1,4}+\sigma _5 \mathfrak{s}_{1,5}+\sigma _6 \mathfrak{s}_{1,6}+\mathfrak{s}_{1,3}=0,\notag\\
 h_2&=\sigma _4 \mathfrak{s}_{1,3,4}+\sigma _5 \mathfrak{s}_{1,3,5}+\sigma _6 \mathfrak{s}_{1,3,6}+\sigma _4
   \sigma _5 \mathfrak{s}_{1,4,5}+\sigma _4 \sigma _6 \mathfrak{s}_{1,4,6}+\sigma _5 \sigma _6
   \mathfrak{s}_{1,5,6}=0,\notag\\
 h_3&=\sigma _4 \sigma _5 \sigma _6 \mathfrak{s}_{2,3}+\sigma _5 \sigma _6 \mathfrak{s}_{2,4}+\sigma _4 \sigma _6
   \mathfrak{s}_{2,5}+\sigma _4 \sigma _5 \mathfrak{s}_{2,6}=0.\notag
\end{align}
The gauge fixed scattering amplitude for scalars is given by
\begin{align}
\label{eq:ampl6sc}
A^{\phi^3}_6=\oint \frac{d\sigma_4d\sigma_5d\sigma_6}{ h_1 h_2 h_3}\frac{\sigma _4 \left(1-\sigma _5\right) \sigma _5 \left(1-\sigma _6\right) \left(\sigma _4-\sigma _6\right) \sigma _6}{\left(1-\sigma _4\right) \left(\sigma _4-\sigma _5\right) \left(\sigma _5-\sigma _6\right)}.
\end{align}
Applying partial fraction decomposition as well as transformations by rational scattering equations (\ref{ratf}), we can rewrite the integrand of (\ref{eq:ampl6sc}) as
\begin{align}
\label{partfrac6}
\frac{\sigma _4 \left(1-\sigma _5\right) \sigma _5 \left(1-\sigma _6\right) \left(\sigma _4-\sigma _6\right) \sigma _6}{\left(1-\sigma _4\right) \left(\sigma _4-\sigma _5\right) \left(\sigma _5-\sigma _6\right)}\hateq \frac{P_1}{\sigma_4-\sigma_5}+P_2
\end{align}
where $P_1$ and $P_2$ are polynomials. To reduce the rational part to a polynomial, we take
\begin{align}
\label{6ptansatz}
	P_1\hateq \left(\sigma _4-\sigma _5\right) P_3
\end{align}
with the following standard form Ansatz:\footnote{Ladder type monomials with base length $m_{max}=n-4$ appear to be a sufficient monomial basis.}
\begin{align}
	P_3=c_1 \sigma _5 \sigma _4^2+c_2 \sigma _4 \sigma _5^2+c_3 \sigma _6 \sigma _4^2+c_4 \sigma _4\sigma _6^2+c_5 \sigma _6 \sigma _5^2+c_6 \sigma _5 \sigma _6^2 +c_7 \sigma _5 \sigma _4+c_8 \sigma _6 \sigma _4+c_9 \sigma _5 \sigma_6.\notag
\end{align}
There are nine constants $c_i$ with $i=1,2,...,9$ we have to fix. We apply the reduction procedure of section \ref{app:degred} to both sides of (\ref{6ptansatz}), collect all terms on one side of the equation and demand that the overall coefficient in front of each monomial vanishes. This produces a set of nine linear equations in nine unknowns. Solving the set of linear equations fixes the nine unknown coefficients and thus yields a polynomial $P_3$. With this, also reducing $P_2$ to contain ladder type monomials only, a standard form numerator polynomial $N^{\phi^3}_6\hateq P_2+P_3$ is obtained. It takes a direct implementation of the polynomial reduction algorithm in \textit{Mathematica} and a linear solver just a few seconds to find a valid analytic $N^{\phi^3}_6$ result, without much effort spent on optimization.\footnote{If we start with the left hand side of eq. (\ref{partfrac6}) instead, as in $\sigma _4 \left(1-\sigma _5\right) \sigma _5 \left(1-\sigma _6\right) \left(\sigma _4-\sigma _6\right) \sigma _6\hateq \left(1-\sigma _4\right) \left(\sigma _4-\sigma _5\right) \left(\sigma _5-\sigma _6\right)N^{\phi^3}_6$, it takes the polynomial reduction algorithm and linear solver, with a few tweaks, about a minute to obtain a different more complicated analytic version of $N^{\phi^3}_6$.} We can evaluate the amplitude making use of prescription (\ref{finalA}):
\begin{align}
\label{eq:ampl6scRes}
A^{\phi^3}_6&=(-1)^3 3!~\text{Res}_{\sigma_{[6}=\infty,~\sigma_{5}=\infty,~\sigma_{4]}=\infty}\left[\frac{N^{\phi^3}_6}{\prod_{j=1}^{n-3} h_j}\right].
\end{align}
The result is completely analytic and about one page long. It can be simplified making use of momentum conservation and on-shell conditions by hand, which is somewhat tedious. Instead we set up a basis of physical poles and fix the coefficients by multiple evaluation on different kinematic points as follows.\\
As in \cite{Huang:2015yka}, the physical poles are given by $\mathfrak{s}_{1,2},\mathfrak{s}_{2,3},\mathfrak{s}_{3,4},\mathfrak{s}_{4,5},\mathfrak{s}_{5,6},\mathfrak{s}_{6,1},\mathfrak{s}_{1,2,3},\mathfrak{s}_{2,3,4}$ and $\mathfrak{s}_{3,4,5}$. By dimensional analysis we see that each term in the amplitude should have three different poles. This means the complete basis is given by $\binom{9}{3}=84$ different triple pole combinations with unknown coefficients. Making use of the procedure described in appendix \ref{app:ratmom}, we can generate $84$ different rational kinematic points and evaluate the amplitude and the basis $84$ times. This sets up a linear set of $84$ equations in the same number of unknowns. Solving this set of equations fixes the coefficients (which turn out to be exactly $1$ or $0$) and yields the simplified $6$-point scalar tree level amplitude in terms of physical poles
\begin{align}
\label{eq:ampl6scRes}
A^{\phi^3}_6=-&\left( \frac{1}{\mathfrak{s}_{1,2} \mathfrak{s}_{3,4} \mathfrak{s}_{5,6}}+\frac{1}{\mathfrak{s}_{1,2} \mathfrak{s}_{5,6} \mathfrak{s}_{1,2,3}}+\frac{1}{\mathfrak{s}_{2,3} \mathfrak{s}_{5,6}
   \mathfrak{s}_{1,2,3}}+\frac{1}{\mathfrak{s}_{1,6} \mathfrak{s}_{2,3} \mathfrak{s}_{2,3,4}}+\frac{1}{\mathfrak{s}_{1,6} \mathfrak{s}_{3,4} \mathfrak{s}_{2,3,4}}\right.\\
	&+\frac{1}{\mathfrak{s}_{2,3}
   \mathfrak{s}_{5,6} \mathfrak{s}_{2,3,4}}+\frac{1}{\mathfrak{s}_{3,4} \mathfrak{s}_{5,6} \mathfrak{s}_{2,3,4}}+\frac{1}{\mathfrak{s}_{1,2} \mathfrak{s}_{3,4}
   \mathfrak{s}_{3,4,5}}+\frac{1}{\mathfrak{s}_{1,6} \mathfrak{s}_{3,4} \mathfrak{s}_{3,4,5}}+\frac{1}{\mathfrak{s}_{1,6} \mathfrak{s}_{2,3} \mathfrak{s}_{4,5}}\notag\\
	&\left.+\frac{1}{\mathfrak{s}_{1,2}
   \mathfrak{s}_{1,2,3} \mathfrak{s}_{4,5}}+\frac{1}{\mathfrak{s}_{2,3} \mathfrak{s}_{1,2,3} \mathfrak{s}_{4,5}}+\frac{1}{\mathfrak{s}_{1,2} \mathfrak{s}_{3,4,5}
   \mathfrak{s}_{4,5}}+\frac{1}{\mathfrak{s}_{1,6} \mathfrak{s}_{3,4,5} \mathfrak{s}_{4,5}}\right),\notag
\end{align}
which is equivalent to summing Feynman diagrams in $\phi^3$ theory and agrees with the result found in \cite{Huang:2015yka}.

				\subsubsection{Six point tree level - first special example}
		Here we will give an example that is very hard to do with less advanced versions of diagrammatic integration rule techniques.\footnote{The author thanks J. Bourjaily for pointing this out and suggesting this test integrand.}  It involves integrating the following terms over the CHY measure
		\begin{align}
\frac{1}{\sigma _{2,3}^4 \sigma _{4,5}^4 \sigma _{6,1}^4}.
\end{align}
Multiplying with the CHY measure and applying our gauge we get
\begin{align}
U_1=\oint \frac{d\sigma_4d\sigma_5d\sigma_6}{ h_1 h_2 h_3}\frac{\left(1-\sigma _4\right) \sigma _4 \left(1-\sigma _5\right) \sigma _5 \left(1-\sigma _6\right) \left(\sigma _4-\sigma _6\right) \left(\sigma _5-\sigma _6\right) \sigma _6}{\left(\sigma _4-\sigma _5\right)^3}.
\end{align}
In order to polynomially reduce the effective rational integrand, we write 
\begin{align}
\label{special1}
\left(1-\sigma _4\right) \sigma _4 \left(1-\sigma _5\right) \sigma _5 \left(1-\sigma _6\right) \left(\sigma _4-\sigma _6\right) \left(\sigma _5-\sigma _6\right) \sigma _6\hateq\left(\sigma _4-\sigma _5\right)^3N
\end{align}
where we use the following standard form polynomial Ansatz
\begin{align}
N=c_1 \sigma _5 \sigma _4^2+c_2 \sigma _4 \sigma _5^2+c_3 \sigma _6 \sigma _4^2+c_4 \sigma _4\sigma _6^2+c_5 \sigma _6 \sigma _5^2+c_6 \sigma _5 \sigma _6^2 +c_7 \sigma _5 \sigma _4+c_8 \sigma _6 \sigma _4+c_9 \sigma _5 \sigma_6.\notag
\end{align}
We have to find nine constants $c_1,c_2,...,c_9$. A completely analytic result is directly accessible applying our procedure, yet not very readable.\footnote{Here an analytic $N$ can be obtained from the polynomial reduction algorithm and a linear solver within $1$ to $2$ minutes. This timing probably could be substantially improved by optimization.} We do not expect the result to be given by pure physical poles either. Therefore, we will instead demonstrate an explicit exact evaluation of the integral on the following kinematic point, which was generated making use of the procedure described in appendix \ref{app:ratmom}:
	\begin{align}
	\label{kin6}
k_1^\mu&=(20,~\,\, 20,~\,\,\,~ 0, \,\,\,\,\,\,\,\, 0),&k_4^\mu&=(\,\,\,\,60, -48,  \,\,\,\,\,~0, -36),\notag\\
k_2^\mu&=(25, -20, \,\,\,\,15, \,\,\,\,\,\,\,\, 0),&k_5^\mu&=(-80, \,\,\,\,48,\,\,\,\, 64, \,\,\,\,\,~0),\\
k_3^\mu&=(39,~\,\,\,~ 0, -15, \,\,\,\,36),&k_6^\mu&=(-64, \,\,\,\,\,~0, -64, \,\,\,\,\,~0).\notag
	\end{align}
First we apply the degree reduction procedure of section \ref{app:degred} to both sides of equation (\ref{special1}) and collect all monomials on one side. The vanishing of the overall coefficient of each monomial separately produces a set of linear equations. Solving this set of equations yields
\begin{align*}
&{\scriptstyle c_5= \frac{7059649218217401322274}{3974168469797996315755} ~~~~\,~~,~~~ c_6= -\frac{5529649875686983344959}{15896673879191985263020} ~~~,~~~ c_2= \frac{12838684423}{1662217245} ~~,~~~ c_4= \frac{354818034905}{57180273228} }\\
& {\scriptstyle c_7=-\frac{5774994253402805042003591}{2146050973690918010507700} ~~,~~~ c_8= -\frac{466431129022169341083793}{343368155790546881681232} ~~,~~~ c_9= -\frac{70384223902707859416469}{158966738791919852630200}~~,~~~c_1=c_3=0.}
	\end{align*}
Using this in the Ansatz for $N$ above, we obtain a standard form numerator polynomial and can apply (\ref{finalA}) to evaluate the integral:
\begin{align}
\label{u1result}
U_1&=(-1)^3 3!~\text{Res}_{\sigma_{[6}=\infty,~\sigma_{5}=\infty,~\sigma_{4]}=\infty}\left[\frac{N}{\prod_{j=1}^{n-3} h_j}\right]\notag\\
&=-\frac{c_2}{\mathfrak{s}_{1,5} \mathfrak{s}_{2,3} \mathfrak{s}_{1,4,5}}+\frac{c_4}{\mathfrak{s}_{1,6} \mathfrak{s}_{2,3} \mathfrak{s}_{1,4,6}}+\frac{c_5}{\mathfrak{s}_{1,5} \mathfrak{s}_{2,3}
   \mathfrak{s}_{1,5,6}}-\frac{c_6}{\mathfrak{s}_{1,6} \mathfrak{s}_{2,3} \mathfrak{s}_{1,5,6}}\notag\\
	&=\frac{14174374134763}{40854136935339786240000}.
\end{align}
Note that indeed properly only the coefficients of highest degree ladder type monomials appear in the final result.\\
Alternatively, we can solve the scattering equations numerically and obtain a numerical approximation for $U_1$, which agrees with (\ref{u1result}).

		\subsubsection{Six point tree level - second special example}
		Another example that is impossible to do with less advanced diagrammatic integration rule techniques involves integrating the following terms over the CHY measure\footnote{Again, the author thanks J. Bourjaily for pointing this out and suggesting this test integrand.}
		\begin{align}
\frac{1}{\sigma _{2,3}^2 \sigma _{3,4}^2 \sigma _{4,2}^2 \sigma _{1,5}^2 \sigma _{5,6}^2 \sigma _{6,1}^2}.
\end{align}
Combining this with the CHY measure and applying our usual gauge we have
\begin{align}
U_2=\oint \frac{d\sigma_4d\sigma_5d\sigma_6}{ h_1 h_2 h_3}\frac{\left(1-\sigma _5\right) \left(\sigma _4-\sigma _5\right) \sigma _5 \left(1-\sigma _6\right) \left(\sigma _4-\sigma _6\right) \sigma _6}{\left(1-\sigma _4\right) \sigma _4 \left(\sigma _5-\sigma _6\right)}.
\end{align}
In order to polynomially reduce the effective rational integrand, we write the equation
\begin{align}
\label{special2}
\left(1-\sigma _5\right) \left(\sigma _4-\sigma _5\right) \sigma _5 \left(1-\sigma _6\right) \left(\sigma _4-\sigma _6\right) \sigma _6\hateq\left(1-\sigma _4\right) \sigma _4 \left(\sigma _5-\sigma _6\right)N
\end{align}
where we use the following standard form polynomial Ansatz
\begin{align}
N=c_1 \sigma _5 \sigma _4^2+c_2 \sigma _4\sigma _5^2 +c_3 \sigma _6 \sigma _4^2+c_4 \sigma _4\sigma _6^2 +c_5 \sigma _6\sigma _5^2 +c_6 \sigma _5 \sigma _6^2+c_7 \sigma _5 \sigma _4+c_8 \sigma _6 \sigma _4+c_9 \sigma _5 \sigma_6.\notag
\end{align}
So that again there are nine constants $c_1,c_2,...,c_9$ to be fixed. Just as before, we can proceed completely analytically, yet the result would be too large to report.\footnote{Here, again, an analytic $N$ can be obtained from the polynomial reduction algorithm and a linear solver within $1$ to $2$ minutes. This timing probably could be substantially improved by optimization.} Therefore, we will illustrate the procedure by evaluating the integral on the kinematic point (\ref{kin6}) instead.\\
First we apply the degree reduction procedure of section \ref{app:degred} to both sides of equation (\ref{special2}) and collect all monomials on one side of the equation. Demanding that the overall coefficient of each monomial vanishes separately provides us with a set of linear equations. Solving the set of equations we obtain
\begin{align*}
 &c_5= \frac{162215379551}{1221259549104} ~~~~~~,~~~ c_6= \frac{5662761717335}{17097633687456} ~~\,~~,~~~ c_4= -\frac{92500}{133623} ~~~,~~~ c_2= \frac{39458}{133623}~~~, \\
& c_7= -\frac{23433636506339}{34195267374912} ~~~,~~\,\, c_8=
   \frac{329688097714075}{273562138999296} ~~~,~~~ c_9= -\frac{3664568494697}{3256692130944},~~~c_1=c_3=0.
	\end{align*}
Plugging this into the Ansatz for $N$ above, we therefore have obtained a standard form numerator polynomial and can use (\ref{finalA}) to evaluate the integral:
\begin{align}
\label{u2result}
U_2&=(-1)^3 3!~\text{Res}_{\sigma_{[6}=\infty,~\sigma_{5}=\infty,~\sigma_{4]}=\infty}\left[\frac{N}{\prod_{j=1}^{n-3} h_j}\right],\notag\\
&=-\frac{c_2}{\mathfrak{s}_{1,5} \mathfrak{s}_{2,3} \mathfrak{s}_{1,4,5}}+\frac{c_4}{\mathfrak{s}_{1,6} \mathfrak{s}_{2,3} \mathfrak{s}_{1,4,6}}+\frac{c_5}{\mathfrak{s}_{1,5} \mathfrak{s}_{2,3}
   \mathfrak{s}_{1,5,6}}-\frac{c_6}{\mathfrak{s}_{1,6} \mathfrak{s}_{2,3} \mathfrak{s}_{1,5,6}}\notag\\
	&=-\frac{2407}{15692753534976}.
\end{align}
Note that again properly only the coefficients of the highest degree ladder type monomials enter the final result. Additionally, it is clear that the calculation for this example structurally follows exactly the same steps and has the same level of complexity as the previous two examples, which would have been different from the point of view of applying diagrammatic integration rules to evaluate the integral.\\
 Alternatively, we can solve the scattering equations numerically and obtain a numerical approximation for $U_2$, which agrees with (\ref{u2result}).

		\subsubsection{Eight point tree level scalar amplitude}
		At eight points there are five scattering equations. The gauge fixed scattering amplitude for scalars reads\footnote{Where $\sigma_3=1$ is implied.}
\begin{align}
\label{eq:ampl8sc}
A^{\phi^3}_8=\oint \frac{\prod_{i=4}^8 d\sigma_i}{\prod_{j=1}^{5} h_j}\frac{\sigma _4 \sigma _5 \sigma _6 \sigma _7 \sigma _8 \sigma _{3,5} \sigma _{3,6} \sigma _{3,7} \sigma _{3,8} \sigma _{4,6} \sigma _{4,7} \sigma _{4,8} \sigma _{5,7} \sigma _{5,8} \sigma _{6,8}}{\sigma _{3,4} \sigma_{4,5} \sigma _{5,6} \sigma _{6,7} \sigma _{7,8}}.
\end{align}
We will demonstrate an explicit evaluation of the amplitude. Making use of the procedure described in appendix \ref{app:ratmom}, we generate some on-shell kinematic data
			\begin{align}
	\label{kin8}
k_1^\mu&=(\,\,\,\,-54, \,\,\,\,-54, ~\,\,\,\,\,\,\,\,0,\,\,\,\,\,\,\,\, 0),&k_5^\mu&=(-85,\,\,\,\, 0,\,\,\,\, 75,\,\,\,\, 40),\notag\\
k_2^\mu&=(-246,\,\,\,\,\,\,\,\, 54, -240, \,\,\,\,\,\,\,\,0),&k_6^\mu&=(\,\,\,\,50,\,\,\,\, 0, -30, -40),\\
k_3^\mu&=(\,\,\,\,260, \,\,\,\,100, \,\,\,\,240, \,\,\,\,\,\,\,\,0),&k_7^\mu&=(-34,\,\,\,\, 0,\,\,\,\, 30, -16),\notag\\
k_4^\mu&=(\,\,\,\,125, -100,\,\,\,\, -75, \,\,\,\,\,\,\,\,0),&k_8^\mu&=(-16,\,\,\,\, 0,\,\,\,\,\,\,\,\, 0,\,\,\,\, 16).\notag
	\end{align}
	We want to find an effective integral expression
\begin{align}
A_8^{\phi^3}=\oint \frac{\prod_{i=4}^8 d\sigma_i}{\prod_{j=1}^{5} h_j}N^{\phi^3}_8,
\end{align}
where $N^{\phi^3}_8$ is a standard form polynomial satisfying
\begin{align}
\label{8ptRedAnsatz}
\sigma _4 \sigma _5 \sigma _6 \sigma _7 \sigma _8 \sigma _{3,5} \sigma _{3,6} \sigma _{3,7} \sigma _{3,8} \sigma _{4,6} \sigma _{4,7} \sigma _{4,8} \sigma _{5,7} \sigma _{5,8} \sigma _{6,8}\hateq \sigma _{3,4} \sigma_{4,5} \sigma _{5,6} \sigma _{6,7} \sigma _{7,8}N^{\phi^3}_8
\end{align}
on the support of the ideal spanned by the scattering equations. As an Ansatz for $N^{\phi^3}_8$ we take the $375$ different ladder type monomials with $m_{max}=n-4=4$. At eight points, polynomially reducing the complete right hand side of (\ref{8ptRedAnsatz}) proves to be time consuming. Therefore, we instead perform a much simpler polynomial reduction of the expression $\sigma_iN_{\sigma_i}\rightarrow N'_{\sigma_i}$ for $i=4,...,8$ with the same Ansatz for $N_{\sigma_i}$.\footnote{The resulting polynomial $N'_{\sigma_i}$ features the same monomials as $N_{\sigma_i}$, but with the coefficients mixed by the reduction procedure.} These results can now be straightforwardly linearly combined as in $(\sigma_i-\sigma_j)N\rightarrow N'_{\sigma_i}-N'_{\sigma_j}\equiv N'_{\sigma_{ij}}$. Additionally, we can nest them by computing the reduction in steps of one degree at a time $(\sigma_i-\sigma_j)(\sigma_a-\sigma_b)N\rightarrow (\sigma_i-\sigma_j)N'_{\sigma_{ab}}\rightarrow N''_{\sigma_{ij}\sigma_{ab}}$, where in the second step we treat the complete monomial coefficients of $N'_{\sigma_{ab}}$ as simple unknowns and substitute their structure back in once the reduction has been performed. Clearly, we can apply the nesting as many times as required. Therefore, the polynomial reduction of $\sigma_iN_{\sigma_i}$ is the only building block we need to construct the complete effective numerator polynomial $N^{\phi^3}_8$.\\
Furthermore, it is more convenient to fractionally decompose the integrand in $(\ref{eq:ampl8sc})$. The numerators and denominators of each of the resulting fractions have smaller polynomial degree, so that the complexity of finding a polynomial reduction for each of these fractions separately is reduced compared to the original expression.\\
Once the polynomial reduction is complete, we collect all terms in (\ref{8ptRedAnsatz}) on one side of the equation and demand the vanishing of all overall monomial coefficients separately. This gives us $375$ linear equations in the same number of unknowns. Solving these equations, we fix the unknown coefficients and obtain the effective standard form numerator polynomial $N^{\phi^3}_8$. With this, prescription (\ref{finalA}) is easily evaluated:
\begin{align}
A_8^{\phi^3}=&(-1)^55!~\text{Res}_{\sigma_{[8}=\infty,~\sigma_{7}=\infty,~\sigma_{6}=\infty,~\sigma_{5}=\infty,~\sigma_{4]}=\infty}\left[\frac{N^{\phi^3}_8}{\prod_{j=1}^{n-3} h_j}\right]=\frac{1360947997721}{22293435818142720000000000000}.\notag
\end{align}
	This is an exact result since we did not invoke any floating point calculations at any step. Alternatively, we can approximately solve the scattering equations numerically and evaluate $A_8^{\phi^3}$ on the solutions, which yields agreement.

\section{CHY formulation of 1-loop level scattering amplitudes}
\label{sec:loop}
At one loop, $n$-point scattering equations have been shown to follow from $(n+2)$-point tree level scattering equations with two massive particles by taking the forward limit of the two massive momenta \cite{He:2015yua}. The tree level scattering equations with two massive particles are given by	\cite{Naculich:2014naa,Naculich:2015zha}:
\begin{align}
\label{massSC}
&~~~E_a=\sum_{{b=1}\atop{b\neq a}}^{n+2}\frac{\mathfrak{p}_{a,b}}{\sigma_{ab}} ~~~\text{for}~~~a\in\{1,2,...,n\},\\
E_{n+1}=\sum_{b=1}^{n}\frac{\mathfrak{p}_{n+1,b}}{\sigma_{n+1,b}} &+\frac{\mathfrak{p}_{n+1,n+2}+m^2}{\sigma_{n+1,n+2}} ~~~,~~~E_{n+2}=\sum_{b=1}^{n}\frac{\mathfrak{p}_{n+2,b}}{\sigma_{n+2,b}} -\frac{\mathfrak{p}_{n+1,n+2}+m^2}{\sigma_{n+1,n+2}}, \notag
\end{align}
where two particles are massive with the same mass $k_{n+1}^2=k_{n+2}^2=m^2$. Here we have introduced a shorthand notation\footnote{When all momenta are massless and on-shell, we have $\mathfrak{p}_{\alpha(1),\alpha(2),...,\alpha(q)}=\mathfrak{s}_{\alpha(1),\alpha(2),...,\alpha(q)}$ from (\ref{kinS}).}
\begin{align}
\label{kinP}
\mathfrak{p}_{\alpha(1),\alpha(2),...,\alpha(q)}\equiv\sum_{\{\beta(1),\beta(2)\}\subset\{\alpha(1),\alpha(2),...,\alpha(q)\}}k_{\beta(1)}\cdot k_{\beta(2)}~~~~~~\text{for integer }q>1.
\end{align}
The sum is over all unordered subsets of two numbers out of a set of $q$ numbers. In the context of 1-loop CHY amplitudes, equations (\ref{massSC}) and (\ref{kinP}) also naturally arise from the formalism described in \cite{Cardona:2016bpi}, without the need to impose them.\footnote{The author thanks C. Cardona and H. Gomez for pointing this out.}\\
In the following we will require the scattering equations in polynomial form. To obtain them, we can for instance apply an appropriate transformation to (\ref{massSC}). However, we should proceed carefully, since in the forward limit 
\begin{align}
k_{n+1}^\mu\to -l^\mu~~~,~~~k_{n+2}^\mu\to l^\mu
\end{align}
the set of equations (\ref{massSC}) admits singular solutions with $\sigma_{ij}\to 0$ for some $i\neq j$, if $E_{n+1}$ and $E_{n+2}$ are taken into consideration. Such singular solutions have no physical contribution to the amplitudes of relevant theories \cite{He:2015yua,Cachazo:2015aol}. Therefore, we will use $(n-1)$ independent equations $E_a$ with $a\leq n$ in order to exclude the singular solutions. It is straightforward to check that the transformation we are looking for is given by
\begin{align}
\label{massSCtrafo}
\tilde{h}_a^{p,q,v} =\sum_{{i=1}\atop{i\neq p,q,v}}^{n+2}\sigma_{ip}\sigma_{iq}\sigma_{iv}Y^{a-2}_{p,q,v,i}E_i~~~\text{for}~~~a\in\{2,3,...,n\},
\end{align}
where
\begin{align}
Y^x_{p,q,v,i}=\left\{\begin{matrix}
         \sum_{\{\alpha(1),...,\alpha(x)\}\subset\{1,...,n+2\}/\{p,q,v,i\}}\prod_{j=1}^x\sigma_{\alpha(j)} & \text{  for }0<x\leq n-2,\\
				 1 & \text{for }x=0,\\
         0 & \text{for $x<0$ and $x>n-2$}.
        \end{matrix}\right.
\end{align}
The range in the index $a$ is set to correspond to (\ref{scatOrig}). Indices $p,q,v$ label the three different massive scattering equations (\ref{massSC}) that are dropped. As we expect, $\tilde{h}_a^{p,n+1,n+2}$ yields the same results regardless of the choice of $p$, so in the following we can consider $\tilde{h}_a^{1,n+1,n+2}$ for convenience. We can compactly write this result as
\begin{align}
\label{massSCpoly}
\tilde{h}_a^{1,n+1,n+2} = \sum_{\{\alpha(1),...,\alpha(a)\}\subset\{1,2,...,n+2\}}(\mathfrak{p}_{\alpha(1),...,\alpha(a)}+m^2\delta_{\alpha,\{n+1,n+2\}})\prod_{j=1}^a \sigma_{\alpha(j)}=0,
\end{align}
for integer $2\leq a \leq n$. Here we used a generalized Kronecker delta
\begin{align}
\delta_{\alpha,\{n+1,n+2\}}=\left\{{{1~~~\text{ if }~~~\{n+1,n+2\}\subset\{\alpha(1),...,\alpha(a)\}}\atop {0~~~\text{ if }~~~\{n+1,n+2\}\nsubset\{\alpha(1),...,\alpha(a)\}}}\right..
\end{align}
As long as we consider $\tilde{h}_a^{1,n+1,n+2}$ in the massive case before taking the forward limit, the scattering equations have the full set of $(n-1)!$ solutions. Knowing that the forward limit is singular in nature, we should check whether any singular solutions resurge in (\ref{massSCpoly}) due to the transformation (\ref{massSCtrafo}) having been applied. Indeed, if we choose to gauge fix $\sigma_1,\sigma_{n+1}$ and $\sigma_{n+2}$, it is straightforward to see that the trivial solution $\sigma_i=\sigma_1$ for $i=2,3,...,n$ is now present in the forward limit,\footnote{Setting $\sigma_i=\sigma_1$ for $i=2,3,...,n$ causes all scattering equations to be proportional to $\mathfrak{p}_{1,2,...,n}$, which vanishes in the forward limit.} additionally to the $(n-1)!-2(n-2)!$ expected regular solutions. Luckily, we can remove this trivial solution by fixing the gauge $\sigma_1=\infty$.\footnote{The fact that the trivial solution can be projected out by a gauge choice indicates that its contribution is not physical.} For convenience we will also fix $\sigma_{n+1}=0,~\sigma_{n+2}=1$. Thus, we will work with the following representation of gauge fixed polynomial scattering equations with two massive particles
\begin{align}
\label{scatL}
 h_i\equiv\left(\lim_{\sigma_1\rightarrow \infty}\frac{1}{\sigma_1}\tilde{h}_{i+1}^{1,n+1,n+2}\right)|_{{\sigma_{n+1}=0}\atop{\sigma_{n+2}=1}}=0~~~,~~~\forall i\in\{1,2,...,n-1\},
\end{align}
which has a smooth forward limit containing only regular solutions of interest.\footnote{We use the same symbol $ h$ as for tree level scattering equations here, since it is always clear from context which scattering equations are in use.} It will be convenient to treat the forward limit as a regulator whenever the kinematics in the limit becomes singular.\\
For $\tilde{h}_a^{1,n+1,n+2}$ the transformation Jacobian is $(-1)^{n+1}[\prod_{i=2}^n\sigma_{1i}][\prod_{1<j<q\leq n+2}^{j\leq n}\sigma_{jq}]$. Therefore, possibly up to a minus sign we have the usual CHY measure for polynomial scattering equations 
\begin{align}
d\mu=\left(\prod_{{c=1}\atop{c\neq q,p,w}}^{n+2} d\sigma_c\right)(\sigma_{qp}\sigma_{pw}\sigma_{wq})\left(\prod_{1\leq i<j\leq n+2}\sigma_{ij}\right)\left(\prod_{a=2}^{n}\delta\left(\tilde{h}_a^{1,n+1,n+2}\right)\right).
\end{align}
Recall that we gauge fixed the moduli $q=1,~p=n+1,~w=n+2$. To test our evaluation procedure at one-loop level, we will consider the bi-adjoint scalar $\phi^3$ theory as proposed in \cite{He:2015yua}, which can be written as
\begin{align}
A_n^{1-loop,\phi^3}=\int\frac{d^Dl}{(2\pi)^D}\frac{1}{l^2}\lim_{{k_{n+1}\to -l}\atop{k_{n+2}\to ~l}}\int d\mu\left(\sum_{\gamma\in\text{cyclic}\{1,2,...,n\}}PT(n+2,\gamma,n+1)\right)^2,
\end{align}
where
\begin{align}
PT(n+2,\gamma,n+1)=\frac{1}{\sigma_{n+2,\gamma(1)}\sigma_{\gamma(1),\gamma(2)}...\sigma_{\gamma(n),n+1}\sigma_{n+1,n+2}}.
\end{align}
However, our evaluation method applies more generally to any integrand that is rational in $\sigma$-moduli and is being integrated over the measure $d\mu$.

\subsection{One-loop amplitude examples}
\label{sec:loopexpl}
		\subsubsection{Two point 1-loop scalar amplitude}
		At two points and 1-loop there is one scattering equation, given by\footnote{Since the forward limit makes the kinematics singular, we use it as a parametrization.}
		\begin{align}
 h_1=\sigma _2 \mathfrak{p}_{1,2}+\mathfrak{p}_{2,3}=0.
\end{align}
		The gauge fixed amplitude amounts to
				\begin{align}
		A_2^{1-loop,\phi^3}=\int\frac{d^Dl}{(2\pi)^D}\frac{1}{l^2}\lim_{{k_{3}\to -l}\atop{k_{4}\to ~l}}\oint  \frac{d\sigma_2}{ h_1}\frac{1}{\left(1-\sigma _2\right) \sigma _2}.
		\end{align}
		We require a standard form numerator polynomial $N_{1-\text{loop}}^{2,\phi^3}$ such that $1\hateq \left(1-\sigma _2\right) \sigma _2 N_{1-\text{loop}}^{2,\phi^3}$ with the standard form Ansatz $N_{1-\text{loop}}^{2,\phi^3}=c_1$. Making use of the scattering equation, we polynomially reduce the right hand side, collect all terms on one side of the equation and in doing so obtain one linear equation in one unknown. Solving this equation and applying momentum conservation yields:
						\begin{align}
N_{1-\text{loop}}^{2,\phi^3}=\frac{\mathfrak{p}_{1,2}^2}{\mathfrak{p}_{2,3}\mathfrak{p}_{2,4}}.
		\end{align}
		Prescription (\ref{finalA}) suggests the calculation
		\begin{align}
		(-1)^1(1!)\text{Res}_{\sigma_2=\infty}\left[\frac{1}{ h_1}\frac{\mathfrak{p}_{1,2}^2}{\mathfrak{p}_{2,3} \mathfrak{p}_{2,4}}\right]=\frac{\mathfrak{p}_{1,2}}{\mathfrak{p}_{2,3} \mathfrak{p}_{2,4}}.
		\end{align}
		If we solve the scattering equation instead $\sigma_2=-\frac{\mathfrak{p}_{2,3}}{\mathfrak{p}_{1,2}}$, we get exactly the same result
				\begin{align}
		\sum_{{ h=0}\atop{\text{solutions}}}\frac{1}{\det\left([\partial_i  h_j]\right)}\frac{1}{\left(1-\sigma _2\right) \sigma _2}=\frac{\mathfrak{p}_{1,2}}{\mathfrak{p}_{2,3} \mathfrak{p}_{2,4}}.
				\end{align}
				In the forward limit we have $\mathfrak{p}_{1,2}\to 0$ while $\mathfrak{p}_{2,3}$ and $\mathfrak{p}_{2,4}$ stay finite. Therefore, the 1-loop integrand vanishes.
				
						\subsubsection{Three point 1-loop scalar amplitude}
		At three points and 1-loop there are two scattering equations, given by
		\begin{align*}
 h_1&=\sigma _2 \mathfrak{p}_{1,2}+\sigma _3 \mathfrak{p}_{1,3}+\mathfrak{p}_{1,5}=0,\\
 h_2&=\sigma _3 \mathfrak{p}_{2,4}+\sigma _2 \mathfrak{p}_{3,4}+\sigma _2 \sigma _3 \mathfrak{p}_{4,5}=0.
\end{align*}
		The gauge fixed amplitude can be written as
		\begin{align*}
		A_3^{1-loop,\phi^3}=\int\frac{d^Dl}{(2\pi)^D}\frac{1}{l^2}\lim_{{k_{4}\to -l}\atop{k_{5}\to ~l}}\oint  \frac{d\sigma_2d\sigma_3}{ h_1 h_2}\frac{-\left(\sigma _2^2+\sigma _3^2-\left(\sigma _2+1\right) \sigma _3\right){}^2}{\left(1-\sigma _2\right) \sigma _2 \left(1-\sigma _3\right)
   \left(\sigma _2-\sigma _3\right) \sigma _3}.
		\end{align*}
		Therefore, we consider the following equality in order to find a standard form effective numerator polynomial $N_{1-\text{loop}}^{3,\phi^3}$
			\begin{align*}
	-\left(\sigma _2^2+\sigma _3^2-\left(\sigma _2+1\right) \sigma _3\right){}^2\hateq \left(1-\sigma _2\right) \sigma _2 \left(1-\sigma _3\right)
   \left(\sigma _2-\sigma _3\right) \sigma _3 N_{1-\text{loop}}^{3,\phi^3},
		\end{align*}
		with the standard form Ansatz $N_{1-\text{loop}}^{3,\phi^3}=c_1 \sigma_2+c_2\sigma_3$. We apply the reduction procedure of section \ref{app:degred} to both sides of this equation, collect all terms on one side and demand that the overall coefficient in front of each monomial vanishes separately. This sets up two linear equations in two unknowns $c_1,~c_2$. Solving for the unknowns yields a numerator polynomial $N_{1-\text{loop}}^{3,\phi^3}$. Using prescription (\ref{finalA}) and simplifying via five-point momentum conservation and on-shell conditions with two massive particles we get the result
 \begin{align}
\label{loopPhi33}
&(-1)^22!\,\text{Res}_{\sigma_{[3}=\infty,~\sigma_{2]}=\infty}\left[\frac{N_{1-\text{loop}}^{3,\phi^3}}{ h_1 h_2}\right]=\\
&=-\frac{1}{\mathfrak{p}_{1,2}}\left(\frac{1}{\mathfrak{p}_{3,5}}+\frac{1}{\mathfrak{p
   }_{3,4}}\right)-\frac{1}{\mathfrak{p}_{2,3}}\left(\frac{1}{\mathfrak{p}_{1,5}}+\frac{1}{\mathfrak{p}_{1,4}}\right)-\frac{1}{\mathfrak{p}_{1,3}}\left(\frac{1}{\mathfrak{p}_{2,5}}+\frac{1}{\mathfrak{p}_{2,4}}\right)-\frac{1}{\mathfrak{p}_{1,5} \mathfrak{p}_{2,4}}-\frac{1}{\mathfrak{p}_{2,5} \mathfrak{p}_{3,4}}-\frac{1}{\mathfrak{p}_{1,4} \mathfrak{p}_{3,5}}.\notag
					\end{align}			
					Alternatively, we can solve the scattering equations and obtain the two solutions $(\sigma_{2,+},\sigma_{3,+})$ and $(\sigma_{2,-},\sigma_{3,-})$ with
					 \begin{align}
					\sigma_{2,\pm}&=\frac{\mathfrak{p}_{1,3} \mathfrak{p}_{3,4}-\mathfrak{p}_{1,5} \mathfrak{p}_{4,5}}{2
   \mathfrak{p}_{1,2} \mathfrak{p}_{4,5}}-\frac{\mathfrak{p}_{2,4}}{2
   \mathfrak{p}_{4,5}}\pm \frac{\sqrt{\left(\mathfrak{p}_{1,2}
   \mathfrak{p}_{2,4}-\mathfrak{p}_{1,3} \mathfrak{p}_{3,4}+\mathfrak{p}_{1,5}
   \mathfrak{p}_{4,5}\right){}^2-4 \mathfrak{p}_{1,2} \mathfrak{p}_{1,5} \mathfrak{p}_{2,4}
   \mathfrak{p}_{4,5}}}{2 \mathfrak{p}_{1,2} \mathfrak{p}_{4,5}},\notag\\
						\sigma_{3,\pm}&=\frac{\mathfrak{p}_{1,2} \mathfrak{p}_{2,4}-\mathfrak{p}_{1,5} \mathfrak{p}_{4,5}}{2
   \mathfrak{p}_{1,3} \mathfrak{p}_{4,5}}-\frac{\mathfrak{p}_{3,4}}{2
   \mathfrak{p}_{4,5}}\mp \frac{\sqrt{\left(\mathfrak{p}_{1,2}
   \mathfrak{p}_{2,4}-\mathfrak{p}_{1,3} \mathfrak{p}_{3,4}+\mathfrak{p}_{1,5}
   \mathfrak{p}_{4,5}\right){}^2-4 \mathfrak{p}_{1,2} \mathfrak{p}_{1,5} \mathfrak{p}_{2,4}
   \mathfrak{p}_{4,5}}}{2 \mathfrak{p}_{1,3} \mathfrak{p}_{4,5}}.\notag
		\end{align}		
		Evaluating the integral on these solutions, summing the contributions and simplifying by means of momentum conservation and on-shell conditions directly leads to exactly the same result (\ref{loopPhi33}).\\
		In the forward limit, terms $\mathfrak{p}_{i,4},\mathfrak{p}_{i,5}$ with $i\in{1,2,3}$ stay finite while $\mathfrak{p}_{i,j}$ with $i,j\in\{1,2,3\}$ tend to zero. Therefore, we first rewrite each of the three different terms in parenthesis in (\ref{loopPhi33}) analogously to the following
			\begin{align}		
     -\frac{1}{\mathfrak{p}_{1,2}} \left(\frac{1}{\mathfrak{p}_{3,5}}+\frac{1}{\mathfrak{p}_{3,4}}\right)=-\frac{1}{\mathfrak{p}_{3,4} \mathfrak{p}_{3,5}}\left(\frac{\mathfrak{p}_{3,4}+\mathfrak{p}_{3,5}}{\mathfrak{p}_{3,4}+\mathfrak{p}_{3,5}+\frac{1}{2}(k_4+k_5)^2}\right).
			\end{align}	
		We may parametrize the forward limit as $k_4^\mu=-(l^\mu+\tau q^\mu_4)$ and $k_5^\mu=(l^\mu+\tau q^\mu_5)$ with $\tau\to0$ and finite $q^\mu_4\neq q^\mu_5$. With this, at leading order we find
		\begin{align}		
    \frac{1}{\mathfrak{p}_{3,l+\tau q_4} \mathfrak{p}_{3,l+\tau q_5}}\left(\frac{\tau\mathfrak{p}_{3,q_5}-\tau\mathfrak{p}_{3,q_4}}{\tau\mathfrak{p}_{3,q_5}-\tau\mathfrak{p}_{3,q_4}+\tau^2\frac{1}{2}(q_5-q_4)^2}\right)=\frac{1}{(\mathfrak{p}_{3,l})^2}+O(\tau).
			\end{align}	
		Therefore, the one-loop integrand at three points in bi-adjoint scalar $\phi^3$ theory is given by
				\begin{align}
		A_3^{1-loop,\phi^3}&=\int\frac{d^Dl}{(2\pi)^D}\frac{1}{l^2}\left( \frac{1}{\mathfrak{p}
   _{1,l} \mathfrak{p}_{2,l}}+\frac{1}{\mathfrak{p}_{1,l}
   \mathfrak{p}_{3,l}}+\frac{1}{\mathfrak{p}_{2,l}
   \mathfrak{p}_{3,l}}+\frac{1}{\mathfrak{p}_{1,l}^2}+\frac{1}{\mathfrak{p}_{2,l}^2}+\frac{1}{\mathfrak{p}_{3,l}^2}\right)\\
	&=\int\frac{d^Dl}{(2\pi)^D}\frac{1}{l^2}\left(\frac{1}{\mathfrak{p}_{1,l}^2}+\frac{1}{\mathfrak{p}_{2,l}^2}+\frac{1}{\mathfrak{p}_{3,l}^2}\right),
			\end{align}	
			since the first three terms vanish by three-point momentum conservation. Since we might be interested in the 1-loop 3-point amplitude as a vertex correction, it would make sense to consider the momenta $k_1,k_2,k_3$ to be off-shell $-$ then the above result is non-trivial. In case when $k_1,k_2,k_3$ are on-shell, all appearing integrals are scaleless.

\section{Conclusion and outlook}
\label{sec:conclusion}
In this work we started with the CHY formulation of scattering amplitudes in arbitrary dimension. We then developed the degree reduction procedure of section \ref{app:degred} and applied it alongside the strong Nullstellensatz to show that any rational function can be written as a standard form polynomial on the support of scattering equations. Making use of this conversion for CHY amplitude integrands, we derived an evaluation prescription that allows to find an amplitude purely from collecting consecutive simple residues at infinity only.\\
Summing over all possible ladder type shapes and taking into account the multiplicity due to available subsets of non-gauge fixed moduli that are used to compose the shapes, we realize that the total number of different ladder type monomials at any $n$ is given by $N^{\text{ladd}}_n=s(n-3)$, where the function $s(x)$ is
\begin{align*}
				s(0)=1,~~~~~s(x)=\sum_{i=0}^{x-1}\binom{x}{i}s(i).
\end{align*}
Upon inspection, the $s(x)$ turn out to be equivalent to so called ordered Bell numbers, or Fubini numbers. For large $x$ these numbers asymptote to $x s(x-1)\approx \ln(2)s(x)$, so that the number of ladder type monomials grows quicker than factorially with $n$.\\
In all explicit amplitude examples we studied above, it was sufficient to consider the subset of ladder type monomials with highest base length $m_{max}=n-4$ to find standard form polynomials corresponding to relevant rational functions. By the counting above, at any $n$ there are $N^{\text{ladd}}_{m_{max}=n-4}=(n-3)s(n-4)$ such ladder type monomials.\\
It is well known that gauge fixed scattering equations have $(n-3)!$ different solutions at tree level \cite{Cachazo:2013gna,Cachazo:2013hca}. In \cite{Cardona:2015ouc,Dolan:2015iln} it was shown that gauge fixed polynomial scattering equations can be transformed to a different form such that $\sigma_i-P_i(\sigma_n)=0$ for $i\in\{4,5,...,n-1\}$ and $P_n(\sigma_n)=0$, where the $P_i(\sigma_n)$ are univariate polynomials in $\sigma_n$. The polynomial $P_n(\sigma_n)$ is of highest degree $(n-3)!$ and accomodates the $(n-3)!$ different solutions. Reducing multivariate polynomials over this transformed system of equations trivially leaves $(n-3)!$ univariate monomials (i.e. $1,\sigma_n,\sigma_n^2,...,\sigma_n^{(n-3)!-1}$) as a minimal basis for the quotient ring of multivariate polynomials over the ideal spanned by scattering equations $Q=R/\langle h_1,h_2,...,h_{n-3}\rangle$. Therefore, the dimension of the quotient ring is $\dim_R(Q)=(n-3)!$ and thus, in the present case, we can similarly expect only $(n-3)!$ of ladder type monomials to be linearly independent on the support of the ideal spanned by scattering equations $\langle h_1,h_2,...,h_{n-3}\rangle$. Here, a natural candidate for such a minimal basis would be the $(n-3)!$ highest degree ladder type monomials. At first glance it might seem that restricting to this minimal basis could increase computational efficiency, since this sets up a minimal linear system of equations in the polynomial construction of rational terms and makes the resulting coefficients unique. However, on a second thought it becomes apparent that modifying the polynomial reduction algorithm such as to eliminate the tail of lower degree ladder type monomials is highly non-trivial and would introduce a large computational overhead before the linear system of equations is set up. Therefore, employing more than the minimal amount of ladder type monomials to keep polynomial reduction simple at the expense of working with larger linear systems of equations appears to be more convenient.
\\
One nice feature of the above procedure is that it works in exactly the same fashion at any $n$ and for amplitudes of any theory in CHY formulation due to the inherent structure of CHY integrands: While the complexity of the kinematic part of a CHY amplitude integrand in a theory like i.e. pure Yang-Mills or gravity is greater compared to massless scalars, the integrand still always is a rational function in the $\sigma$-moduli, such that the conceptual steps towards finding the amplitude described in previous sections still remain exactly the same, making the procedure universal. Furthermore, since all relevant residues for any amplitude or partial term in consideration are always collected from simple poles at infinity only, each generic evaluation step is of low complexity and the difficulty is shifted towards finding standard form polynomial expressions for the originally rational amplitude integrands. The polynomial reduction procedure that addresses this problem can be implemented algorithmically in general, so that the amplitude evaluation becomes automated for general input, which is one further strength of the current approach.\\
One problem that is bound to appear as we choose higher values for $n$, is the question of efficiency. The number of linear equations and corresponding number of unknowns increases as $(n-3)s(n-4)$ if we apply the construction step of section \ref{sec:polinv}. Even though other techniques to find the reduced form might exist, this kind of limitation is bound to appear whenever a solution is formulated algorithmically involving a sequence of structural steps leading from a certain input to an output of a different structure. Therefore, as a possible direction for further investigation it might be interesting to search for general $n$-point integrands of standard polynomial form in various theories of interest directly, eliminating the necessity for the polynomial reduction procedure. Additionally, knowing that only the highest degree ladder type monomials contribute to any integral, finding just the coefficients for the minimal basis of highest degree ladder type monomials based on some general physical arguments would be equivalent to obtaining a direct closed form expression for the amplitude, since the remaining contour integration is trivial.

\appendix

\section{Generating real rational on-shell momenta}
\label{app:ratmom}
Pythagorean triples are integers $a,b,c$ such that the relation $c^2=a^2+b^2$ is satisfied. The following well known parametrization of all such triples due to Euclid is convenient
\begin{align}
\label{euclid}
a=h(u^2-v^2)~~~~~~,~~~~~~b=2h u v~~~~~~,~~~~~~c=h(u^2+v^2),
\end{align}
where $h,u,v$ are arbitrary integers. Thinking of an $n$-point amplitude, we can consider $n-2$ separate copies of these integers $a_i,b_i,c_i,h_i,u_i,v_i$ with $i\in\{1,2,...,n-2\}$. We would like to use the above to parametrize $n$ massless external momenta obeying momentum conservation. For that end, we distribute the integers $a_i,b_i,c_i$ into Minkowski momenta components in a fashion similar to the following.
\begin{align}
1)&\text{ Fill $a_1$ into $k^0_1$ (with a random overall sign $\pm$ in front) and $k^1_1$ components, such that: }\notag\\
&~~~k^{\mu}_1=(\pm a_1,a_1,0,...,0)\notag
\end{align}
\begin{align}
2)&\text{ Fill $a_i,b_i$ into spatial components and $\pm c_i$ (random sign) into the zero component }\notag\\
& \text{ of vectors $k^\mu_j$ for $j\in\{2,3,...,n-1\}$ so that $a_q$ and $b_{q+1}$ always appear in consecutive}\notag\\
& \text{ vectors and in the same spatial component but with opposite sign, such that i.e.:}\notag\\
&~~~k^{\mu}_2=(\pm c_1,-a_1,\,\,\,\,b_1,~\,\,\,\,\,0,~\,0,...,0)\notag\\
&~~~k^{\mu}_3=(\pm c_2,~\,\,\,\,\,0,-a_2,\,\,\,\,b_2,~\,0,...,0)\notag\\
&~~~k^{\mu}_4=(\pm c_3,~\,\,\,\,\,0,~\,\,\,\,\,0,-a_3,b_3,...,0)\notag\\
&~~~\vdots\notag\\
&~~~k^{\mu}_{n-2}=(\pm c_{n-3},0,...,-a_{n-3},\,\,\,\,\,b_{n-3},~~\,\,\,\,0)\notag\\
&~~~k^{\mu}_{n-1}=(\pm c_{n-2},0,...,~~\,\,~\,\,\,\,\,0,-a_{n-2},b_{n-2})\notag
\end{align}
\begin{align}
3)&\text{ Fill $b_{n-2}$ into $k^0_n$ and $k^i_n$ components, pairing the spatial component of $k^i_{n-1}$, i.e.: }\notag\\
&~~~k^{\mu}_n=(\pm b_{n-2},0,...,0,0,-b_{n-2})\notag
\end{align}
Since each set of $a_i,b_i,c_i$ integers is internally parametrized by (\ref{euclid}), all momenta defined above are automatically light-like $k_i\cdot k_i=0$ for $i\in\{1,2,...,n\}$. Furthermore, if we ensure that $b_q=a_{q+1}$ for all $q\in\{1,2,...,n-3\}$, then all spatial components will sum up to zero, providing spatial momentum conservation. The set of constraints $b_q=a_{q+1}$ can be solved using $n-3$ of the $h_i$ of (\ref{euclid}) and promoting them to variables. Finally, to ensure momentum conservation in the zero-th component, we can solve the equation $\sum_{i=1}^nk_i^0=0$ in $u_1$ of (\ref{euclid}) while promoting it to a variable. The solutions to the constraints above are rational in the unfixed parameters, so that we are guaranteed to obtain rational momenta if we seed integers to the unfixed $h_{n-2}$ and $u_i,v_i$. However, we should seed the integers carefully since singular configurations exist. In order to avoid most singular results we could for instance fix $u_i=1$ for all remaining $i$, while randomly selecting $h_{n-2},v_i>1$. Finally, it is clear that the position of the spatial components within a vector can be assigned flexibly as long as the canceling entries, such as $b_q$ and $-a_{q+1}$, always are properly paired. Therefore, we can randomly create real rational on-shell momenta in any spacetime dimension $D>2$ using the above. Even though this only provides access to a very specific subset of all possible real and rational on-shell momenta, they are nevertheless sufficiently generic for testing purposes. Straightforward slight modifications can also be made to obtain sufficiently generic results even for the four point configuration, or cases involving massive particles.

\acknowledgments
The author would like to thank A. Volovich for initial collaboration on this paper, especially for pointing out the relevance of the global residue theorem and counting of monomials. The author also thanks M. Spradlin and A. Volovich for reading the draft and useful comments and discussions.  The author thanks J. Bourjaily for interesting discussions on the comparison with other integration techniques. This work is supported by the US Department of Energy under contract DE-FG02-11ER41742.

%
%



%

\end{document}